\documentclass[aps,prx,reprint]{revtex4-2}

\usepackage{blindtext}
\usepackage{graphicx}
\usepackage{dcolumn}
\usepackage{bm}
\usepackage{mathtools, nccmath}
\usepackage{hyperref}
\usepackage{siunitx}
\UseRawInputEncoding

\begin{document}
\title{Unraveling and controlling the self-assembly pathways of cubic colloids}

\author{Dillip Kumar Mohapatra\textsuperscript{\dag}{*}}
\author{Teun W.J. Verouden\textsuperscript{\dag}}
\author{Janne-Mieke Meijer\textsuperscript{*}}

\affiliation{Department of Applied Physics and Science Education, Eindhoven University of Technology, Groene Loper 19, 5612 AP Eindhoven, The Netherlands.}
\affiliation{Institute for Complex Molecular Systems, Eindhoven University of Technology, Groene Loper 19, 5612 AP Eindhoven, The Netherlands.}

\altaffiliation{\textsuperscript{\dag}These authors contributed equally to this work.\textsuperscript{*}Corresponding author.}
\email{d.k.mohapatra@tue.nl, j.m.meijer@tue.nl}

\begin{abstract}
The self-assembly of anisotropic building blocks into complex spatial architectures is an important design strategy in material science but the mechanisms by which the anisotropic interactions influence the early-stage growth and formation of disordered (non-)equilibrium structures remain poorly understood. Here, we experimentally demonstrate that tuning the strength of shape-induced directional bonds changes the self-assembly pathways of cubic colloids. By tracking the growth kinetics and internal reorganizations of small clusters at increasing attraction strength, we identify three self-assembly regimes: (i) nucleation and growth regime: slow reorganization-dominated growth of crystalline clusters, (ii) dynamic regime: diffusion-limited growth with dynamic cube reorganizations leading to disordered crystalline clusters and (iii) static regime: diffusion-limited growth of kinetically arrested clusters unable to reorganize due to directional bonding constraints. We further show that transitions between these regimes are reversible and allow pathway engineering to control the structure and disorder. Our results reveal how directional bonding governs pathway selection, providing important insights for the rational design of reconfigurable colloidal, nano‑, and biomaterials.
\end{abstract}

\maketitle

\section{Introduction}

The directional interactions between nanoscale and microscale building blocks are key to the formation of (hierarchical) superstructures with exceptional functionality, far beyond the chemical and physical properties of their constituent particles \cite{ whitesides_self-assembly_2002,manoharan_colloidal_2015,boles_self-assembly_2016}. 
Examples include functional materials self-assembled from nano- and micro-particles \cite{li_colloidal_2022},  DNA(-mediated) nanostructures\cite{Zhou_2026_DNA,Zhou_2024_DNAparticles} to biomolecular condensates and protein crystallization \cite{mcmanus_physics_2016}. 
Breakthroughs in synthesis have provided nano- and micro-particles with a wide variety of complex shapes and directional interactions rivaling their atomic and molecular counterparts in complexity \cite{hueckel_total_2021,Kim_2023_adv}. 
Numerous studies have focused on understanding the relationship between the individual particle properties and the superstructures that can form \cite{ damasceno_predictive_2012, boles_self-assembly_2016, li_assembly_2016, Torquato_2018, dijkstra_predictive_2021}. 
These studies have shown that particle alignment and directional interactions dictate the ordered (crystalline) superstructures that can contain remarkable complexity, including quasicrystals \cite{haji-akbari_disordered_2009,noya_2021} and diamond lattices \cite{he_colloidal_2020,Neophytou_2021}.
The reported ordered superstructures, however, are mostly equilibrium structures, and it remains unclear how the self-assembly pathway influences the nucleation and growth kinetics that cause the formation of disordered (non-)equilibrium structures. Especially, in crystalline structures of nano- and microparticles with complex shapes and interactions, abundant crystal defects have been found \cite{li_colloidal_2011, bodnarchuk_structural_2011, von_freymann_bottom-up_2013, serafin_frustrated_2021}. 
Moreover, strong kinetic trapping could result in disordered hierarchical networks of arrested material due to gelation (or aggregation) as known for complex (bio)molecules \cite{cardinaux_interplay_2007, fusco_soft_2016, mcmanus_physics_2016} and microparticles \cite{royall_real_2021}. 

Cubic particles are anisotropic building blocks of particular interest, and have been studied
extensively on both the nano- and micron-scale \cite{henzie_self-assembly_2012,zhao_entropic_2011,rossi_shape-sensitive_2015, meijer_observation_2017, geuchies_situ_2016,lu_2019_SciAdv}.
It has been shown that the exact details of cube shape and the directionality of the interparticle interactions result in a variety of equilibrium superstructures in both two dimensions (2D) and three dimensions (3D), including rotator crystals \cite{batten_phase_2010, ni_phase_2012, meijer_observation_2017,avendano_phase_2012}, tetratic phases \cite{avendano_phase_2012, zhao_entropic_2011, loffler_tetratic_2025}, simple cubic crystals \cite{rossi_cubic_2011, henzie_self-assembly_2012}, rhombic crystals \cite{rossi_shape-sensitive_2015,meijer_observation_2017} and even checker-board structures \cite{wang_self-assembly_2024}. 
However, experimental studies have also reported several different types of disorder and defects in the crystalline structures, including structural polymorphs \cite{quan_solvent-mediated_2014, brunner_selfassembled_2017, meijer_convectively_2019}, grain boundaries \cite{cimada_dasilva_fundamental_2021} or (high) vacancy concentrations \cite{van_der_burgt_cuboidal_2018, smallenburg_vacancy-stabilized_2012}, and that experimental conditions can overrule otherwise dominant interparticle interactions, such as shape and magnetic dipole decoupling in confinement\cite{Baldauf_2022}. 
To achieve high-quality crystals, experimentalists often rely on extensive empirical sampling to determine the protocol with a desired structural outcome \cite{cimada_dasilva_fundamental_2021}, highlighting the importance of the self-assembly pathway. 
To understand the formation of these disordered structures, spatiotemporal information of the anisotropic self-assembly processes is needed but is experimentally challenging. Recent studies on nanoparticle assemblies with liquid-phase transmission electron microscopy (TEM) have provided unique insight into overall crystallization processes \cite{zhong_engineering_2024, luo_unravelling_2023}. However, so far the details of how the strength of directional bonding controls the competition between aggregation, internal rearrangement, and kinetic arrest during the earliest stages of nucleation and growth remain lacking. 

Here, we investigate the nucleation and growth of clusters of micron-sized cubic colloids and show how the strength of the directional interactions influences the self-assembly process. We achieve real-time, particle-resolved observation of the nucleation, growth and structural evolution of small clusters by combining high-speed confocal laser scanning microscopy (CLSM) with a temperature tunable attraction based on critical Casimir forces \cite{hertlein_direct_2008,gambassi_critical_2024}. The critical Casimir force provides reversible \textit{in-situ} control over short-range attractions allowing for in- and out-of-equilibrium assembly of colloidal particles to be induced, as shown for spheres \cite{nguyen_critical_2016, rouwhorst_nonequilibrium_2020}, patchy particles \cite{ swinkels_visualizing_2023, swinkels_networks_2024} and also cubic colloids \cite{kennedy_self-assembly_2022}. 
The short-range attraction combined with the cube shape results in directional interactions whose interaction strength can be precisely tuned, enabling systematic investigations of the growth kinetics and internal cluster reorganizations, involving distinct sliding and turning of cubes due to geometrical constrains. We identify three distinct assembly regimes: (i) nucleation and growth regime at weak attractions yielding slow reorganization-dominated growth of clusters; (ii) dynamic regime at intermediate attractions with diffusion-limited growth with dynamic cube reorganization; and (iii) static regime at strong attractions where diffusion-limited growth occurs without reorganizations due to kinetic trapping and directional bonds. Moreover, we apply these mechanistic insights and show how transitions between the regimes facilitate pathway engineering and control over the superstructure outcome. 
Our results provide direct experimental insight into how the self-assembly pathway of colloids with directional interactions controls the emergence of disorder and (non)-equilibrium superstructures and offers a framework for engineering complex superstructures with desired properties in colloidal, nanoscale and biomolecular systems. 

\section{Results}

\begin{figure*}[hbt!]
    \centering
    \includegraphics[width=0.9\textwidth]{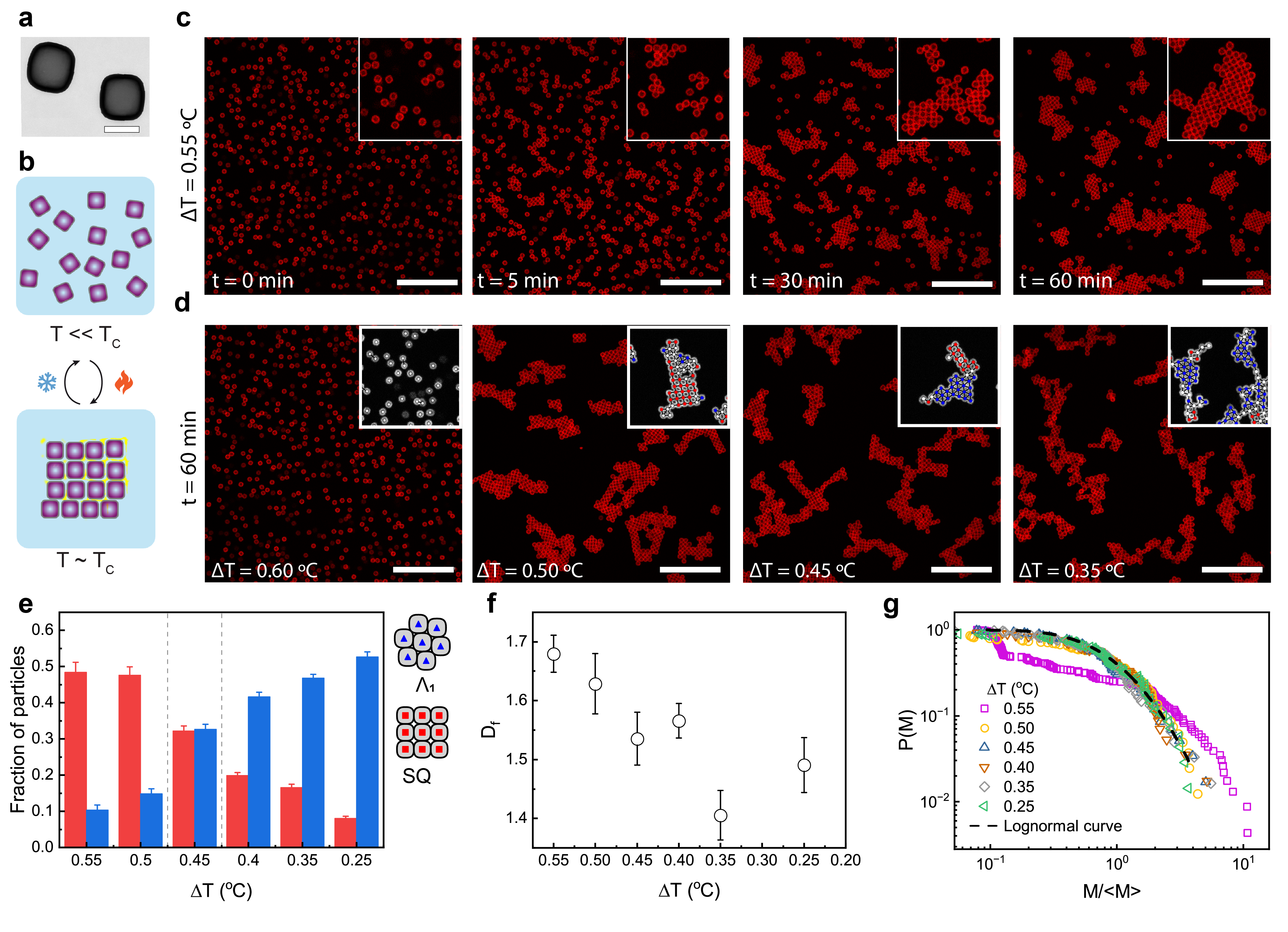}
    \caption{\textbf{The self-assembly of cubic colloids into different clusters at different attraction strength.} 
    (a) Transmission electron microscopy image of hollow silica cubes.  
    (b) Schematic illustration of the critical Casimir attraction between colloidal cubes in a lutidine water mixture controlled by the temperature difference with the critical point $\Delta T = T_C - T$ that leads to different attraction strengths. 
    (c) CLSM images obtained at different times during the self-assembly process of colloidal cube dispersions with an area fraction of $\phi_A=0.21$ at \(\Delta T = 0.55~^\circ\mathrm{C}\). (Insets) higher magnification of the observed cube clusters.
    (d) CLSM images of the final colloidal cube superstructures after 60 min obtained at the other investigated \(\Delta T\). (Insets) Cube clusters with detected particle positions color-coded according to their local arrangement (red) Square-lattice, (blue) \(\Lambda_1\)-lattice and (white) unassigned. 
    (e-g) Structural analysis of final colloidal cube cluster superstructures for all investigated \(\Delta T\). (e) Fraction of particles assigned Square-lattice or \(\Lambda_1\)-lattice, (f) Fractal dimension of the clusters, and (g) normalized complementary cumulative cluster size distribution. Black dashed line is the log-normal cluster size distribution expected for DLCA. Scale bars are (a) $1~ \mu \mathrm{m} $ (c) and (d) $20~ \mu \mathrm{m} $ .}
    \label{fig:structure_evolution1}
\end{figure*}

\textbf{Cube superstructures}. To investigate the self-assembly process of cubic colloids, we employ fluorescently labeled hollow silica cubes \cite{rossi_cubic_2011,meijer_observation_2017} with edge length $L = 1.40\pm 0.07 ~\mu\mathrm{m}$ and distinct rounded corners, as reflected in a shape parameter $m = 2.9$  (Fig.~\ref{fig:structure_evolution1}a). To induce and control the strength of the attractive interactions between the hydrophilic silica cubes, we disperse the cubic colloids in a binary solvent mixture of lutidine in water ($28.4 ~\mathrm{wt}\%$) that phase separates above a critical temperature, $T_{c}$. For sample temperatures of $T \ll T_{c}$ the cubes interact repulsively due to the presence of charges on their surface and remain stable in dispersion. Raising the sample temperature $T$ within  \( \Delta T \approx0.5~^\circ\mathrm{C}\) of $T_{c}$ results in attractive interactions between the cubes, as observed previously\cite{kennedy_self-assembly_2022} (Fig. \ref{fig:structure_evolution1}b). The strength of the attractive interactions induced by the critical lutidine-water mixture can be controlled from a few $k_B T$ to tens of $k_B T$ within this temperature range \cite{hertlein_direct_2008}. The increase in strength is also accompanied by an increase in interaction range from 10 - 100 nm but compared the micron-sized cubes employed here the interactions can be considered short-ranged at all temperatures. All sample temperatures are reported as \(\Delta T = T_c - T\), with lower $\Delta T$ indicating a higher attraction strength and longer attraction range. To control the sample temperature and visualize the positions of the cubes over time, we employed confocal laser scanning microscopy equipped with a temperature-controlled set-up that provide \(\pm0.02~^\circ\mathrm{C}\) temperature stability (see Methods and Fig. S1 for experimental setup details). 

The self-assembly process of the cubic colloids was investigated by performing several temperature controlled experiments in a sample with an area fraction of cubes of $\phi_A = 0.21$. The sample is first equilibrated at $1~^0\mathrm{C}$ below $T_{c}$, during which the cubes sediment onto the capillary wall due to their size and density difference with the solvent. Next, while imaging the system, the temperature is raised to the desired $\Delta T$ to start the self-assembly process and the system is imaged for a duration of 60 min. Figure \ref{fig:structure_evolution1}c and movie S1 show the time evolution of the self-assembly process at \(\Delta T = 0.55~^\circ\mathrm{C}\). During the self-assembly process attractive interactions between the cubes and the hydrophilic silica surface of the capillary wall are also involved and can be observed to restrict the motion of the cubes to 2D. Next, small cube nuclei form and grow gradually via single-particle addition and cluster–cluster attachment, while at the same time detachment and cube reorganization events occur. Finally, large crystalline cube clusters are observed that coexist with a dilute phase of single cubes that confirms that the attractive interactions between the cubes are on the order of $\approx k_BT$ for this $\Delta T$. 

We investigated the influence of the attraction strength on the self-assembled cubic colloid clusters, by repeating the self-assembly experiment at several different temperatures; \(\Delta T = 0.60, 0.50\), \(0.45\), \(0.40\), \(0.35\) and \(0.25~^\circ\mathrm{C}\). In addition, we detected and analyzed all the cube positions within the clusters and classify their local packing arrangement. For these rounded cubic colloids interacting via short-range attractions a switch from a square (SQ) lattice to a rhombic lattice, or $\Lambda_1$-lattice \cite{jiao_optimal_2008}, is expected \cite{kennedy_self-assembly_2022, rossi_shape-sensitive_2015,zhong_engineering_2024} (See methods, supplementary note 1, Fig. S2 and Fig. S3 for structure assignment criteria). We find a distinct difference in the final cube cluster structure obtained at the end of the self-assembly process for decreasing $\Delta T$ (Fig \ref{fig:structure_evolution1}d).  At \(\Delta T = 0.60~^\circ\mathrm{C}\) no significant attraction is present as no clusters are formed. For \(\Delta T \leq 0.55~^\circ\mathrm{C}\) and thus increasing attraction strength, we observe that cube clusters are present and that they show a set of structural transitions in morphology and packing, ranging from small dense clusters with local SQ lattices to large clusters with distinct branches with a local rhombic $\Lambda_1$ lattice, resulting in a mixed phase at $\Delta T = 0.45~^\circ \mathrm{C}$ (insets Fig \ref{fig:structure_evolution1}d and Fig~\ref{fig:structure_evolution1}e). These final structure fractions appear to have reached steady state (see Supplementary Fig. S4). Our results show that under similar self-assembly conditions, i.e. particle area fraction and duration of the process, the attraction strength influences not only the local packing but also the morphology of the cubic colloid clusters.

To quantify the morphology of the cubic colloid clusters, we analyzed the fractal dimension, \(D_f\), and the size distribution, $P(M)$, of the clusters. We determined $D_f$ from the scaling relation between the number of particles in a cluster \(N\) and the radius of gyration of the cluster \(R_g\), given by \(R_g \sim N^{1/D_f}\) (Fig. \ref{fig:structure_evolution1}f). For spherical colloids with attractive interactions, both reaction-limited cluster aggregation (RLCA) or diffusion-limited cluster aggregation (DLCA) can result in fractal clusters. RLCA and DLCA are typically distinguished based on the $D_f$ and size distributions of the clusters and for DLCA at long times and/or higher $\phi_A$ a percolating (gel) network is expected to form \cite{lin_universality_1989,lu_gelation_2008, rouwhorst_nonequilibrium_2020}. We find that the cube clusters exhibit a $D_f = 1.7$ for weak attractions that decreases to that expected for RLCA (\(D_f=1.55\)) at $\Delta T <0.45~^\circ\mathrm{C} $ and finally that of DLCA (\(D_f=1.45\)) at $\Delta T <0.35~^\circ\mathrm{C}$ \cite{robinson_experimental_1992, moncho2001dlca}. The normalized complementary cumulative size distribution rescaled by the mean cluster size $\langle M \rangle$, $P(M)$ (Fig. \ref{fig:structure_evolution1}f) shows that most cluster size distributions collapse on a single log-normal curve for \(\Delta T \leq0.50~^\circ\mathrm{C}\), while \(\Delta T =0.55~^\circ\mathrm{C}\) deviates significantly. Interestingly, this would mean agreement with DLCA occurs at almost all attraction strengths \cite{friedlander1966self} and that agreement with RLCA occurs for the weakest attraction strength only.
Although $D_f$ values might differ for cubes as recently shown in simulations of small anisotropic microparticle aggregates \cite{Rusen2023}, the $D_f$ and $P(M)$ values at different attraction strengths are not in mutual agreement for either RLCA or DLCA. Therefore, we conclude that the cube shape impacts the self-assembly process and cluster growth kinetics.

\textbf{Self-assembly pathway}

\begin{figure*}[hbt!]
    \centering
    \includegraphics[width=0.9\textwidth]{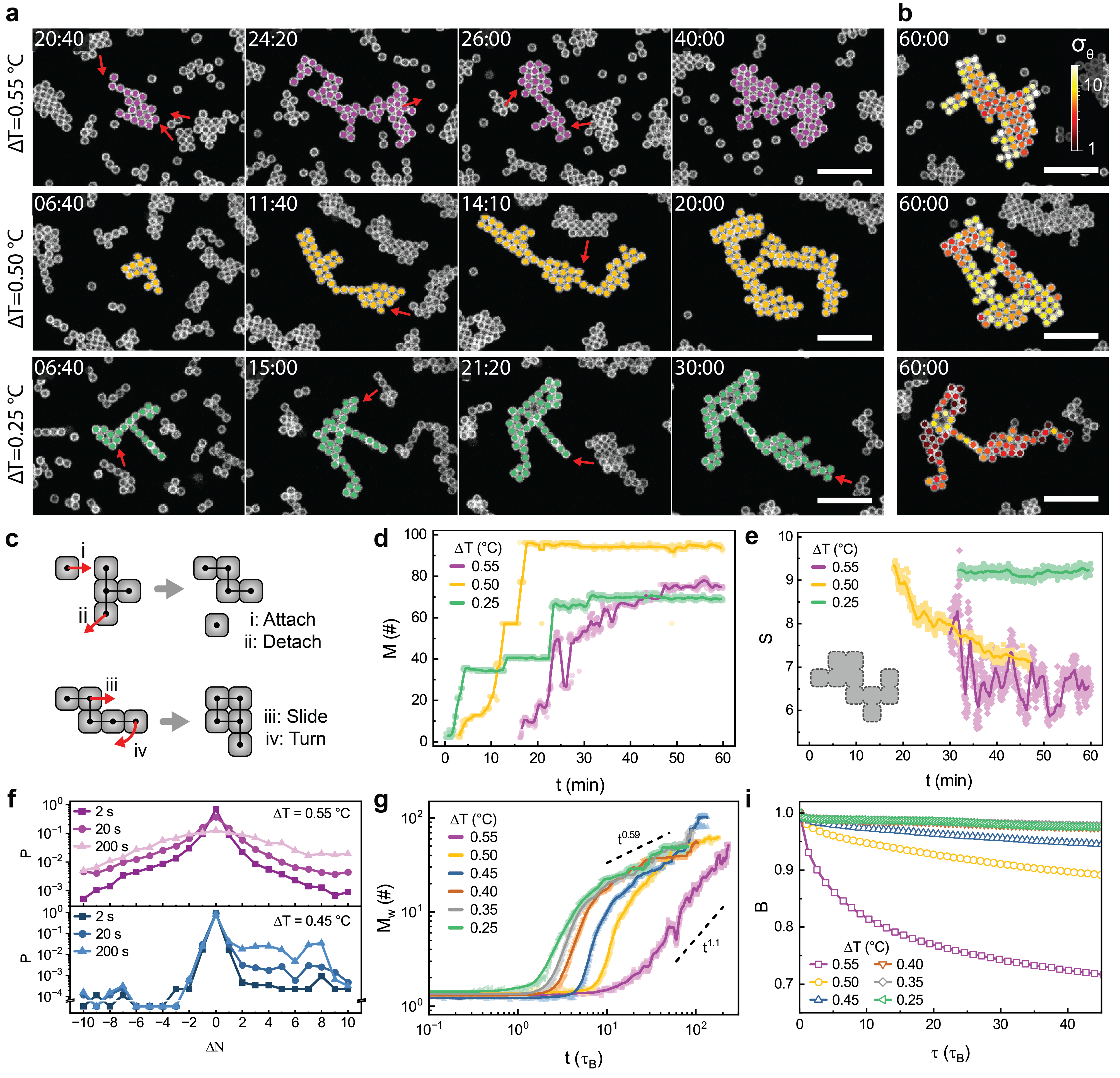}
    \caption{\textbf{The growth and dynamics of cubic colloid clusters.} 
    (a) CLSM images of single cube cluster over time with detected particle positions for (top row, purple) \(\Delta T = 0.55~^\circ\mathrm{C}\), (middle row, yellow) \(\Delta T = 0.50~^\circ\mathrm{C}\), and  (bottom row, green) \(\Delta T = 0.25~^\circ\mathrm{C}\).  (b) Clusters at $t = 60 $ min with detected particles positions colored according to bond angle fluctuations. (c) Schematic representation of observed cube dynamics, (i) attachment, (ii) detachment, (iii) sliding and (iv) turning. (d) Cluster size and (e) Shape parameter of each cluster in (a-b) over time. (f) Bulk growth and shrinking probability after a certain $\Delta t$ for clusters forming at (top, putple) \(\Delta T=0.55~^\circ\mathrm{C}\) and (bottom, blue) \(\Delta T = 0.45~^\circ\mathrm{C}\). (g) Weighted average cluster size and (i) mean bond correlation $B$ for all investigated \(\Delta T\) as a function of Brownian time $\tau_B$.}
    \label{fig:cluster_growth}
\end{figure*}

To unravel how the shape-induced directional interactions of cubic colloids and attraction strength influence the self-assembly process, we investigated the nucleation and growth of cube clusters in more detail. Figure ~\ref{fig:cluster_growth}a (and Movies S2, S3 and S4) shows CLSM images and detected particle positions at different time points for single clusters forming at $\Delta T = 0.55, ~0.50$ and $0.25~ ^\circ \mathrm{C}$ that result in a dense SQ cluster, a slightly branched SQ cluster, and a branched $\Lambda_1$ cluster, respectively. The images reveal that besides different growth rates of the clusters, also distinct internal cluster dynamics are present or absent. The internal cluster dynamics are visualized by the change in bond angle fluctuations over time, $\sigma_\theta$ (Fig. ~\ref{fig:cluster_growth}b) and show that at weak attraction dynamics are present while at strong attraction the dynamics are absent, indicating kinetically arrested clusters.
We identify that during the self-assembly of the cubic colloids two key processes can be distinguished: cluster growth and cube reorganization dynamics (Fig. ~\ref{fig:cluster_growth}c). The cluster growth occurs via the (i) attachment of single cubes (or clusters) and (ii) detachment of cubes and the cube reorganization dynamics occur via two distinct motions (iii) sliding of a cube (or lines of cubes) along other cube faces or (iv) a cube turning around another cube corner. A similar observation of sliding rows of cubes were recently made for cubic nanoparticles \cite{luo_unravelling_2023}. While the growth process is similar to spherical colloids, the presence of two distinct reorganization processes is a result of the geometry of the rounded cubic shape.

To determine when each process occurs, we investigated the growth and morphology of the three clusters over time. Figure ~\ref{fig:cluster_growth}d) shows that the cluster size, $M$, fluctuates over time for the lowest attraction strength but increases step-wise for the two higher attraction strengths. The difference indicates that both attachment and detachment of single cubes or sub-clusters can occur at weak attraction but that at higher attraction strengths initial rapid cube attachment is followed by slower cluster-cluster aggregation and cube detachment is almost absent. To characterize the cube reorganization dynamics we analyzed the overall cluster morphology via a shape parameter \(S = \frac{c}{\sqrt{A}}\), where $c$ is the circumference and $A$ the area of a cluster (Fig.~\ref{fig:cluster_growth}e). For the two lowest attraction strengths, $S$ decreases over time even when for $\Delta T =0.50~^\circ \mathrm{C}$ the cluster size remains constant, while for the highest attraction strength, $S$ is constant over time. For the latter the absence of reorganizations is clearly due to the directional bonding constraints, preventing cubes to turn around a corner. Our analysis shows that growth process changes, i.e. detachment of cubes stops, and cube reorganization dynamics changes, i.e. sliding and turning stop, both cease with increasing attraction strength but do so at different attraction strengths.

To confirm the growth process change with increasing attraction strength, we analyzed the growth and particle dynamics of all clusters observed within the field of view for all investigated $\Delta T$. We determine the attraction strength where cube detachment ceases by analyzing the bulk growth and shrinking probability $P_b(\Delta N)$, which give the change that a cluster gains $\Delta N$ particles in a given time interval (Fig.~\ref{fig:cluster_growth}f and Supplementary Fig. S5).
When the distribution of $P_b(\Delta N)$ is symmetric, an equal probability between attachment ($\Delta N >0$) and detachment ($\Delta N <0$) is evident, while an asymmetric distribution reveals a decrease in detachment \cite{ginot_aggregation-fragmentation_2018, rouwhorst_nonequilibrium_2020}. We find that $P_b(\Delta N)$ is nearly symmetric for $\Delta T = 0.55~ ^\circ \mathrm{C} $, while for $\Delta T \leq 0.50~^\circ \mathrm{C}$ the distribution becomes asymmetric, as most clearly seen for $\Delta T = 0.45 ~^\circ \mathrm{C}$. Fig.~\ref{fig:cluster_growth}g shows the weighted average cluster size over time, given by $M_w(t) = \sum_i i^2n_i(t)/\sum_i in_i(t)$, where $n_i(t)$ is the number of clusters consisting of $i$ particles at time $t$. Here time is rescaled with the measured Brownian time, $\tau_B$, to correct for the slower diffusion of the cubes when they are attracted to the capillary wall (See Supplementary Note 2, Fig. S6 and Fig. S7). For the higher attraction strengths ($\Delta T \leq0.50~^\circ \mathrm{C}$) cluster growth occurs rapidly, indicating quick particle–particle attachment at early times, with the onset shifting to progressively shorter times as the attraction increases in strength. At longer times, the growth slows down due to slower cluster–cluster aggregation limited by the reduced diffusivity of large clusters, although the limited field-of-view of the microscope might also play a role. The mean cluster size growth at long times follows with a power-law growth with an exponent $\alpha = 0.59$ in agreement with that for DLCA kinetics \cite{moncho2001dlca}. For the lowest attraction strength ($\Delta T = 0.55 ~^\circ \mathrm{C}$), the onset of cluster growth is delayed by an order of magnitude in time and we find the mean cluster size to grow instead following a power law with $\alpha = 1.1$ that qualitatively matches the expected value for RLCA \cite{moncho2001dlca}. The overall growth kinetics of the cubic colloids clusters confirm the crossover from cube detachment, or reaction-limited growth, for the weakest attraction strength to the absence of cube detachment, or diffusion-limited growth, for the other attraction strengths. 

Next, we assessed at which attraction strengths the internal cluster reorganizations are present and absent for all investigated $\Delta T$. For this we determining the bond correlation function, $B$, of bonds within clusters over time, given by: 
$B(\tau) = \sum_{\substack{i, j = 1 \\ i \neq j}}^{N} \langle b_{ij}(t) b_{ij}(t+\tau) \rangle$
where $N$ is the number of particles, $b_{ij}(t)$ is the bond between particle $i$ and $j$ at reference time $t$ and $b_{ij}(t+\tau)$ is the same bond a time interval $\tau$ later. We note that decorrelation of \(B(\tau)\) occurs when a bond is broken and therefore captures both detachment and reorganizations events. However, by combining $B(\tau)$ with the growth probability $P(\Delta N)$, the presence of internal cluster reorganization and cube dynamics can be determined. Figure~\ref{fig:cluster_growth}i shows that for the lowest attraction strength \(B(\tau)\) decays rapidly, reflecting the frequent breaking and reforming of bonds, while for the highest attraction strength \(B(\tau)\) is almost constant, indicating the absence of any bond breaking events. Interestingly, for the two intermediate attraction strengths ($\Delta T =0.50, ~0.45~^\circ \mathrm{C}$) the decay in \(B(\tau)\) reveals that particle dynamics are still present while $P(\Delta N)$ showed cube detachment is almost absent. Clearly, internal cluster reorganizations due to the cubes dynamics are present at several attractions strength and only cease for the higher attraction strengths. This finding explains why the cubic colloid cluster's $D_F$ remains high at weak attraction strengths and  highlights the important role of the anisotropic shape of the cubes, as the sliding of cubes remains possible at attraction strengths $> k_B T$.

\begin{figure*}[ht!]
    \centering\includegraphics[width=0.8\textwidth]{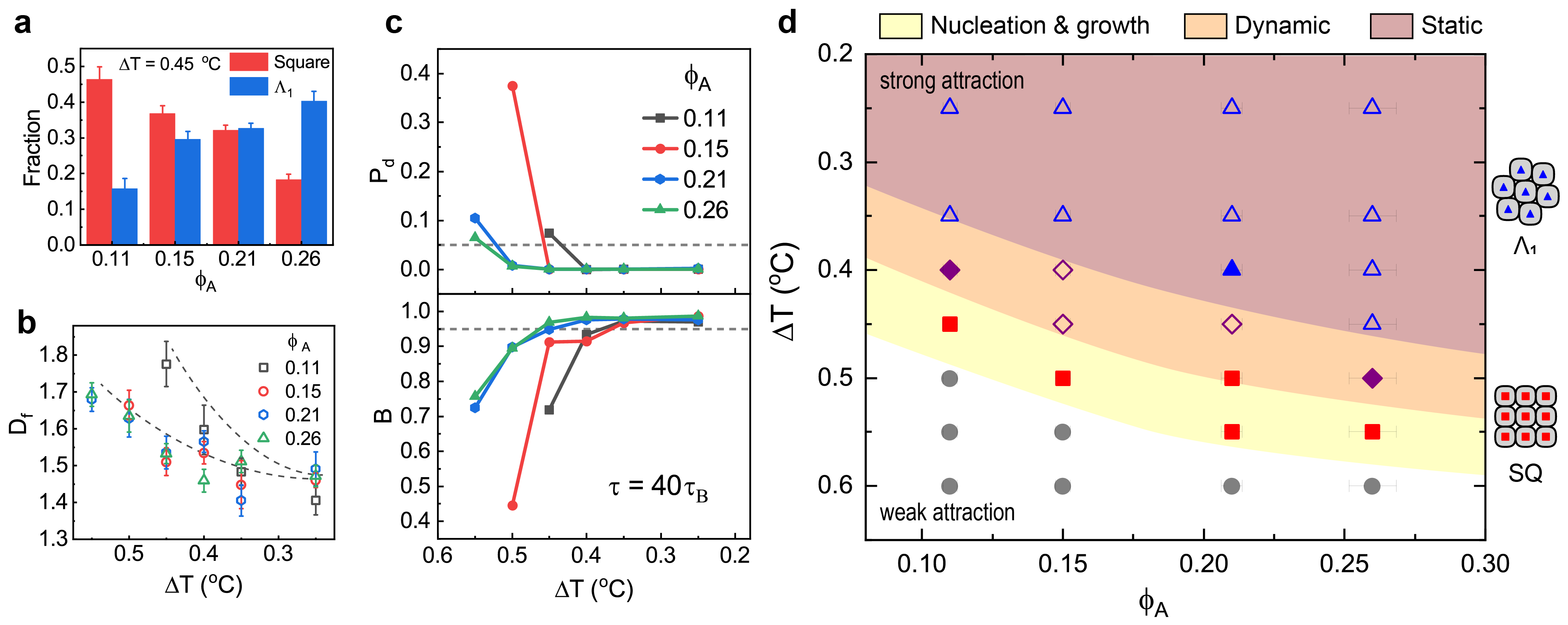}
    \caption{\textbf{Three self-assembly regimes identified by the analysis of the growth process and cube dynamics}. (a) The observed local packing of the cube clusters at \(\Delta T\) =\(0.45~^\circ \mathrm{C}\) after the full self-assembly process, with increasing particle area fraction, $\phi_A$ the lattice structure in the clusters switches from square, SQ, to rhombic, $\Lambda_1$. (b) Fractal dimension of cube clusters $D_f$ that decreases with \(\Delta T\) for all \(\phi_A\) but at at different $\Delta T$ that directly correlates to the attraction strength. (c) Bond break probability $P_b$ and bond correlation $B$ at $\tau = 40 \tau_B$ for different \(\phi_A\) showing that where cube detachment stops $P_b(\tau)<0.05$ and where cube dynamics cease $B(\tau)>0.95$ occur at different $\Delta T$. (d) Experimental state diagram established for the cubic colloids. Note that the $\Delta T$ scale on the $y$-axis decreases from bottom to top to reflect increasing attraction. Where the transition from an isotropic fluid (grey circles) to dense SQ-lattice clusters (red squares) and branched $\Lambda_1$ clusters (blue triangle) occurs via a mixed phase (purple diamonds). Closed symbols represent fluid or dense clusters with \(D_f > 1.55\) and open symbols represent fractal clusters where \(D_f < 1.55\). The shaded areas indicate the different observed regimes: nucleation and growth (yellow), dynamic (orange), and static (dark red). The overall regions are guides to the eye.}
    \label{fig:phase_diagram}
\end{figure*}

Our observations reveal that the growth kinetics and cluster dynamics are controlled by  the directional interactions and the strength of the short-ranged attractive interactions between the cubic colloids. To capture the complexity to the self-assembly process, we distinguish three self-assembly regimes: (1) \textit{nucleation and growth regime} at weak attraction of $\approx k_BT$, where cube attachment and detachment both occur, cubes reorganize frequently and crystalline SQ clusters grow with kinetics characteristic of RLCA; (2) \textit{dynamic regime} at intermediate attraction $> k_BT$, where cube detachment ceases but cube reorganizations do not, resulting in branched SQ clusters with growth behavior characteristic of DLCA; (3) \textit{static regime} at strong attraction $\gg k_BT$, where particle detachments appear fully absent, cube reorganizations do not seem to occur and the formation of elongated fractal $\Lambda_1$ clusters is in line with DLCA characteristics.

\textbf{State diagram}

Next, we investigated the influence of the cube concentration on the 2D self-assembly and growth process of the cubic colloid clusters. Concentration is well known to influence the phase behavior of colloids in general \cite{Fluids_Colloids_and_Soft_Materials.ch12} and that of superballs in particular \cite{batten_phase_2010,meijer_observation_2017,ni_phase_2012}. For three additional particle area fractions of $\phi_A = 0.11,0.15$, and $0.26$ we analyzed the cluster growth, morphology and dynamics at each $\Delta T$ employed above. Overall, we observe similar self-assembly behavior of the cubic colloids for all $\phi_A$; upon a decrease in $\Delta T$, and thus increasing interaction strength, first compact SQ clusters form, followed by branched clusters with mixed SQ and $\Lambda_1$ symmetry, and finally fractal $\Lambda_1$ clusters at the highest attraction strength (See Supplementary Fig. S8). We also find that as $\phi_A$ increases, the onset of aggregation shifts to weaker attraction and cluster growth occurs at higher $\Delta T$. The shift in onset changes the boundaries of the three regimes we identified, as reflected in the transition from SQ to $\Lambda_1$ clusters with increasing $\phi_A$ when $\Delta T$ is constant (Fig.~\ref{fig:phase_diagram}a) and the observation of higher $D_f$ for the same $\Delta T$ for lower $\phi_A$ (Fig.~\ref{fig:phase_diagram}b). 
We determined for each $\phi_A$ where detachment and reorganizations stop, by examining at $\tau = 40 \tau_B$ when the bond break probability decreases to $P_d(\tau)<0.05$ and when the bond correlation exceeds $B(\tau)>0.95$  (Fig.~\ref{fig:phase_diagram}c). These criteria can be used to distinguish the three different self-assembly regimes for all concentrations.

Based on our analysis of the structure and dynamics of all self-assembled cube clusters at the investigated $\Delta T$ and $\phi_A$, we construct a state diagram (Fig.~\ref{fig:phase_diagram} d). We note that the $\Delta T$ scale is inverted to show increasing attractive interactions from bottom to top. The state diagram reveals that the \textit{nucleation and growth regime}, \textit{dynamic regime}, and \textit{static regime} are all present and that $\Delta T$ and $\phi_A$ both influence their boundaries. For an increase in the particle concentration $\phi_A$ the onset of aggregation occurs at higher $\Delta T$ and is related to an increase in total free energy contribution and a decrease in translational entropy as expected \cite{Fluids_Colloids_and_Soft_Materials.ch12}, and also observed for patchy particles with critical Casimir interactions \cite{swinkels_phases_2023}. The $\Delta T$ range over which the regimes are present, however, seem to remain similar in size. We conclude that the increase in the strength and range of the attractive interactions between these cubic colloids determines the structural switch from dense SQ clusters to branched $\Lambda_1$ clusters, from RLCA to DLCA growth, and where cube dynamics freeze. Clearly, our study shows that the presence and absence of cube reorganization dynamics is finely tuned by the interaction strength of the directional interactions and thus must be a key parameter in controlling the self-assembly pathway and structural outcome.

\textbf{Pathway engineering}

\begin{figure*}[ht!]
    \centering
    \includegraphics[width=0.9\textwidth]{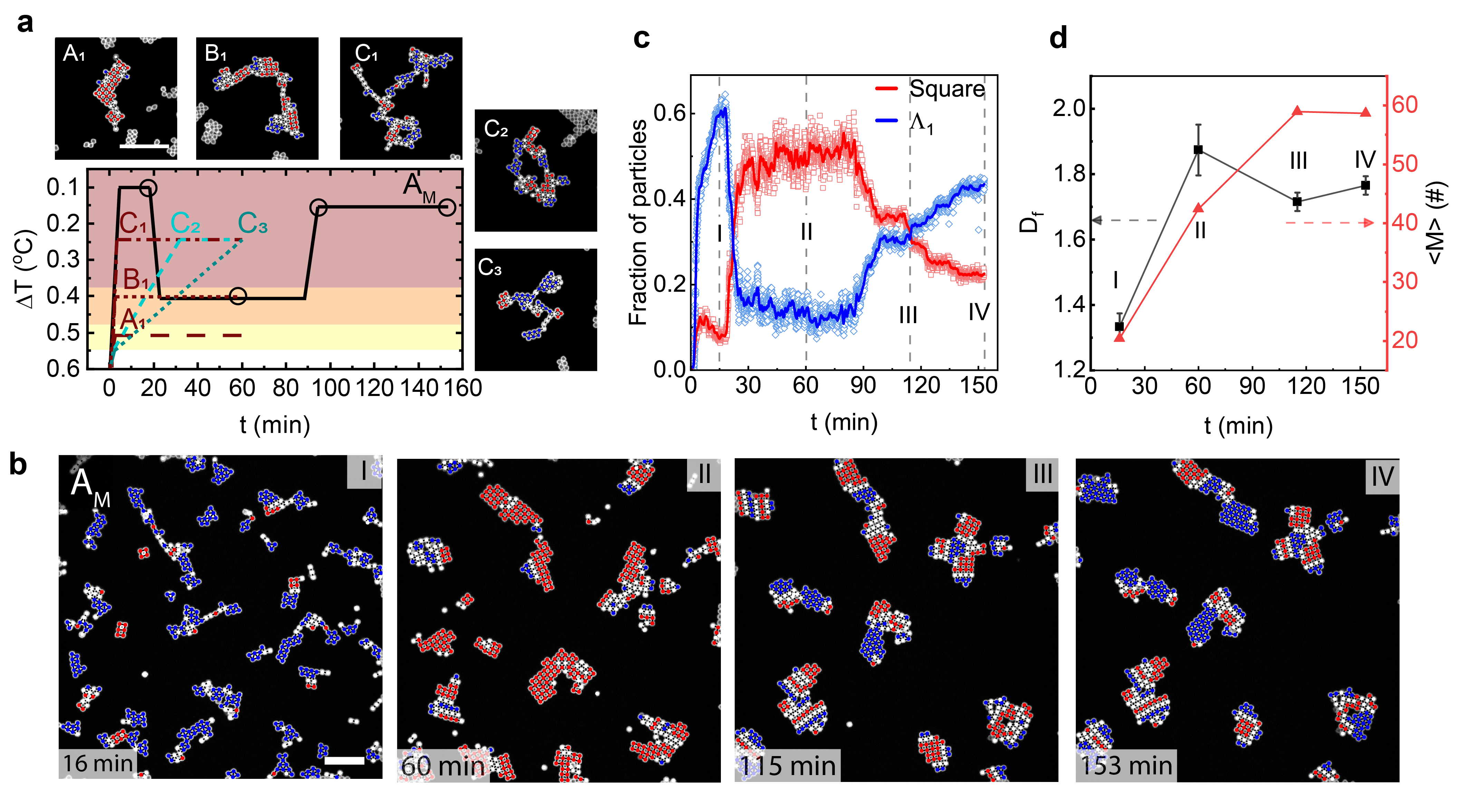}
    \caption{\textbf{The self-assembly pathway controls the superstructure morphology and size.}
    (a) Different thermal pathways applied to a cube dispersion at \(\phi_A = 0.15\) with corresponding CLSM images of single clusters with detected particle positions colour-coded according to their local arrangement (red) Square-lattice, (blue) \(\Lambda_1\)-lattice and (white) unassigned. 
    (b) Time-lapse confocal images along \(A_M\) showing reversible transitions between branched \(\Lambda_1\), square, and dense \(\Lambda_1\) states at 16 (I), 60 (II), 115 (III), and 153 min (IV) (\(\phi_A = 0.15\)). Particles are colour-coded by local structure.
    (c) Temporal evolution of square and \(\Lambda_1\) fractions during the modulated cycle (\(A_M\)).
    (d) Corresponding fractal dimensions and mean cluster sizes at time points I - IV in \(A_M\).}
    \label{fig:Influence_pathways}
\end{figure*}

While the state diagram provides clear insights into the structures that are expected to form at a specific attraction strength, the question remains how a modulation of the strength of the short-range attraction during the self-assembly process could influence the superstructure formation. We therefore investigated how different self-assembly pathways, set by controlled changes in $\Delta T$ over time, influence the cluster structure and formation. Figure ~\ref{fig:Influence_pathways}a shows six different temperature pathways that were applied to cube dispersion at \(\phi_A = 0.15\), together with CLSM images of single clusters with structural assignment. All pathways start from $\Delta T=0.6~^\circ\mathrm{C}$, where no aggregation is expected, and can be classified into three types: a temperature jump (\(A_1, B_1, C_1\)) at $t=0$ minutes, a slow temperature ramp (\(C_2, C_3\)) and a modulated pathway ($A_M$), where the latter two transition between the \textit{nucleation and growth}, the \textit{dynamic} and the \textit{static} regimes in different ways. The resulting clusters structures show that pathways $A_1$, $B_1$ and $C_1$ lead to the expected dense SQ clusters, branched mixed SQ and $\Lambda_1$ clusters and fractal \(\Lambda_1\) clusters at the respective final $\Delta T$. In contrast, for pathways \(C_2\) and \(C_3\) that end at the same final $\Delta T$ as $C_1$, we find that when cluster formation starts in the \textit{nucleation and growth regime} and is followed by growth in the \textit{dynamic regime} before reaching the \textit{static regime}, denser clusters with overall more SQ packing are obtained (see Supplementary Fig. S9). Clearly, the duration in a specific regime during the self-assembly process influences the cluster structure and morphology. 

Finally, we engineered the pathway $A_M$ such that cluster nucleation would occur in the \textit{static regime}, followed by growth and structural reorganization in the\textit{ dynamic regime} and finally return to the \textit{static regime}. Figure ~\ref{fig:Influence_pathways}b and supplementary Movie S5 show the observed clusters and structural assignment during the $A_M$ pathway over time. Here, we indeed observe rapid formation of static fractal \(\Lambda_1\) clusters that upon lowering the attraction strength show distinct reorganizations via sliding and turning of cubes into dense SQ clusters. Finally, upon returning to strong attraction, we observe that the ordering in the clusters transitions from SQ \(\rightarrow \Lambda_1\). The temporal evolution of the local packing, fractal dimension and cluster size confirms the reversibility between the $\Lambda_1 \leftrightarrow$ SQ structures and show that the clusters continuously grow and reorganize into denser structures (Fig.~\ref{fig:Influence_pathways}(c,d)). Interestingly, the transition from $\Lambda_1 \rightarrow$ SQ  appears almost instantaneous and both fractions quickly reach a plateau value. In contrast, the reverse transition SQ \(\rightarrow \Lambda_1\) is initially quite rapid but slows down considerably. 
The continued transition is slightly unexpected, since we previously found reorganizations to be absent at this $\Delta T$. Upon closer inspection we find that rows of SQ cubes slide along each other in the clusters until they form $\Lambda_1$, as evident by unassigned rows of cubes between SQ and $\Lambda_1$ rows (Fig.~\ref{fig:Influence_pathways}b). Combined with the earlier observation that the fractal $\Lambda_1$ clusters consist mostly of branches of single cubes and reorganization would thus require cube turning, we conclude that cube turning stops earlier at strong attraction. A similar observation was made in colloidal clusters (joints) of cubes and spheres where turning around a cube corner is absent \cite{chakraborty_colloidal_2017, shelke_flexible_2023}. Clearly, by tuning the interaction strength in the $A_M$ pathway different self-assembly regime can be accessed, resulting in larger and denser $\Lambda_1$ clusters compared to a temperature jump at the same final $\Delta T$. 

\section{Discussion}
In summary, our experimental results show that the self-assembly pathways of attractive cubic colloids are governed by the strength of the cubes' directional interactions that dictates the growth kinetics and cube reorganization dynamics. We identified three distinct self-assembly regimes with increasing attraction strength:a nucleation and growth regime, a dynamic regime and a static regime, each with different growth kinetics and cube dynamics. 

Whereas part of our observations for the self-assembly process of the micron-sized cubic colloids are consistent with two recent studies on cubic nanoparticle crystallization \cite{luo_unravelling_2023,zhong_engineering_2024}, our study also reveals how the formation of disordered structures occurs due to directional bonding and kinetic trapping. Specifically, the observation of a \textit{nucleation and growth regime} defined by weak, short-range attractions between cubes that results in dense SQ clusters is in agreement with the observed crystallization pathways of nanocubes into SQ (simple cubic) lattices \cite{luo_unravelling_2023,zhong_engineering_2024} and the morphology of the SQ clusters can be explained by thermal roughening of the crystal surface at weak interaction \cite{luo_unravelling_2023}. In contrast, the observation of the \textit{dynamic regime} and the \textit{static regime} characterized by intermediate to strong short-range interactions that result in fractal cluster growth and kinetic trapping and the important role of cube reorganization dynamics was not made in either study. Interestingly, the many observations of defects and disorder in other cubic nanoparticle crystals \cite{quan_solvent-mediated_2014, brunner_selfassembled_2017, cimada_dasilva_fundamental_2021, van_der_burgt_cuboidal_2018} are strong indications that a dynamic and static self-assembly regime must also exist for nanoparticles. 

We further showed that by modulating the attraction strength over time the self-assembly pathway can be engineered by accessing specific growth and dynamics regimes and that the final superstructures can be controlled. Our observation of a direct link between interaction strength and the presence or absence of cube reorganization dynamics provide new insight into self-assembly protocols. For instance, we can compare our pathway engineering result with the extensive empirical study of daSilva \textit{et al.}\cite{cimada_dasilva_fundamental_2021}, who showed that to achieve monocrystalline $\Lambda_1$ assemblies of nanocubes (although not labeled such) the choice of anti-solvent, prevention of solvent contamination and ligand destabilization are crucial. Based on our findings, their optimum protocol must involve a self-assembly pathway with a prolonged duration in the dynamic self-assembly regime where defects and disorder are allowed to anneal out. To gain more insight into the incorporation of defects and their dynamics within crystal lattices of cubic colloids, further studies at higher area fractions of cubic colloids should be conducted. In addition, the influence of the range of the attractive interactions, which can be long range for nanoparticles, might change the cube reorganization dynamics and should be investigated further. 

Our results provide direct insight into the mechanisms that dictate the anisotropic particle self-assembly pathways and reveals that the directional interaction strength is a key parameter. Our approach combined with the abundance of colloidal shapes available today \cite{hueckel_total_2021}, opens up the possibility to address the role of anisotropic interactions and interaction strength modulation during crystallization, where shape has been predicted to matter \cite{gartner_design_2024}, but also gelation, where studies of anisotropic colloids remain limited \cite{royall_real_2021}. 
The mechanistic insights from our study provide a general framework for the rational design and programming of functional hierarchical structures in colloidal, nano-, and bio-materials. 

\section{Methods}

\textbf{Colloidal cube dispersion.}
Hollow silica cubes were prepared using a template-based synthesis in which cubic hematite particles served as sacrificial templates. The process began with the synthesis of micrometer-sized hematite cubes using the gel–sol method proposed by Sugimoto \cite{sugimoto_preparation_1992}. These hematite cubes were subsequently coated with an amorphous silica layer via the Stöber method \cite{rossi_cubic_2011,graf_general_2003,meijer_colloidal_2015}. For confocal laser scanning microscopy, the silica shells were functionalized with rhodamine dye. To obtain hollow silica cubes, the hematite cores were removed by dispersing the silica-coated hematite cubes in concentrated hydrochloric acid (HCl). The resulting structures, often referred to as \emph{superballs}, can be described by the equation

\begin{equation}
  \left|\frac{x}{a}\right|^{m} + \left|\frac{y}{a}\right|^{m} + \left|\frac{z}{a}\right|^{m} \leq 1
\end{equation}

where, $m$ is the shape parameter, here $m=$ 2.9.
The hollow silica cubes are stored in ethanol, as they gradually degrade when dispersed in water. For sample preparation, the dispersion was centrifuged, and the sedimented cubes were redispersed and washed three times with a binary solvent mixture of 2,6-lutidine and water (28.4~wt\%) to ensure complete removal of ethanol. Finally, an appropriate amount of the lutidine/water mixture was added to achieve the desired particle concentration for the experiment.
The cube dispersions were placed in small capillaries (Borosilicate glass, 0.2 x 2 x 50 mm, Vitrocom) and sealed with teflon grease on both ends. 

\textbf{Experimental methods.}
To visualize and control the self-assembly of colloidal cubes, we used a Nikon AR1-25HD confocal laser scanning microscope equipped with a custom-built temperature-controlled setup capable of maintaining a stability of $0.02~^\circ\mathrm{C}$ (Fig.~S1, Supplemental). The microscope equipped with a 100$\times$ oil immersion objective (Nikon CFI Plan Apo VC, NA = 1.4) and a 561 nm laser for excitation of the rhodamine-functionalized silica shells, together with a GaAsP PMT detector. 
Samples were equilibrated at $T = 33~^\circ\mathrm{C}$ prior to imaging. Next, the temperature was rapidly increased to the desired temperature to trigger the self-assembly process with increasing interaction strength. Time-series of 2D images were recorded over an 82 $\times$ 82 \unit{\micro\meter} field of view at a resolution of 1024 $\times$ 1024 pixels with a frame rate of 0.5 frame/s for 60 min. After each measurement, the sample temperature was reduced to $33~^\circ\mathrm{C}$ and held for 10~min to allow the system to return to its initial dispersed state, after which the next measurement was performed.
After completing all measurements, the system was further heated to determine the solvent critical temperature, $T_{c}$, corresponding to phase separation of the lutidine/water mixture. For the concentration used here, $T_{c} = 34~^\circ\mathrm{C}$. Because $T_{c}$ is highly sensitive to solvent composition, it must be determined in situ. All temperatures are therefore reported relative to the critical temperature as $\Delta T = T_{c} - T$.

\textbf{Image analysis.}
For each frame, the centers of the hollow particles were found by performing a circle Hough transform, and locating the resulting bright spots using Trackpy \cite{allan_daniel_b_2021_4682814}. 
The neighboring particles were determined with a cut-off distance of $1.76$ \unit{\micro\meter} (21 pixels), which is approximately $1.25$ times the diameter of the particle, so large enough that all six nearest neighbors are counted in $\lambda_1$ structures, but small enough that the diagonals in square structures are not counted.
These Nearest-Neighbors (NNs) were used to determine both the structural environment of a particle and to identify clusters.
To determine the structural environment of a particle, 2D bond order parameters, which check for $n$-fold symmetry are used:

\begin{equation}
    \psi_n \left( j \right) = \frac{1}{N_b}\sum \limits_{k=1}^{N_b} e^{i n\phi_{jk}}
    \label{eq:2BOP}
\end{equation}

In this analysis, \( N_b \) represents the number of nearest neighbors of particle \( j \), and \( \phi_{jk} \) is the polar angle of the bond between particles \( j \) and \( k \) with respect to the horizontal axis. Particles with \( \psi_4 > 0.85 \) were classified as "square", while those with \( \psi_6 > 0.85 \) were classified as \( \Lambda_1 \) (more information on the distinction between \( \Lambda_0 \) and \( \Lambda_1 \) can be found in the supplementary information section). If both values were either smaller or larger than \( 0.85 \), the structure was classified as unassigned, similar to the approach used by Kennedy \textit{et al.} \cite{kennedy_self-assembly_2022}.
The dominant structure at a given temperature was determined by comparing the relative fractions of each structure type: if the fraction \(X\%\) of one structure was more than twice that of the next most prevalent structure (\(X\% > 2Y\%\)), structure \(X\) was classified as dominant. The final structure and morphology of the clusters were obtained by averaging the bond order parameters and the fractal dimension over the last 300 frames of each measurement, since the recorded field of view contained only a limited number of clusters.
The particle clusters were determined based on the nearest-neighbor distance using Freud \cite{freud2020}. For all subsequent analysis, such as cluster size, structure, and radius of gyration, the particle centers were used, and the particles were assigned equal mass.

\section*{References}

\bibliography{manuscript_cube_assembly}

\section*{Acknowledgements}
The authors thank Paul P.A.M. van der Schoot and Liesbeth M. C. Janssen for insightful discussions and critical reading of the manuscript. This work was supported by the Netherlands Organization for Scientific Research (NWO) grant VI.Vidi.223.126.  

\section*{Author contributions}

\noindent Conceptualization: J.M.M. 

\noindent Methodology: D.K.M., T.W.J.V. and J.M.M. 

\noindent Investigation: D.K.M. and T.W.J.V.  

\noindent Data Analysis: D.K.M., T.W.J.V. and J.M.M. 

\noindent Writing - original draft: D.K.M., T.W.J.V., and J.M.M. 

\section*{Competing interests}
The authors declare no conflict of interest. 

\section*{Data availability}
Data is available from the authors upon reasonable request.

\end{document}